**ZnO/ZnS heterostructures as hole reservoir to boost Ni foam energy storage performance**

Alessia Fischetti[1,2], Giacometta Mineo[1*], Daniela Russo[1,2], Francesco Salutari[3], Claudio Lentini Campallegio[1,2], Elena Bruno[1,2], Jordi Arbiol[3,4], Giorgia Franzò[2], Salvatore Mirabella[1,2], Vincenzina Strano[2*] and M. Chiara Spadaro[1,2,3*]

[1] Physics and Astronomy Department "Ettore Majorana", University of Catania, via S. Sofia 64, Catania 95123, Italy

[2] IMM-CNR, Via S. Sofia 64, 95123 Catania, Italy

[3] Catalan Institute of Nanoscience and Nanotechnology (ICN2), CSIC and BIST, Campus UAB, Barcelona, 08193, Spain

[4] ICREA, Pg. Lluís Companys 23, Barcelona, Catalonia, 08010, Spain

*corresponding authors:

giacometta.mineo@unict.it;

vincenzina.strano@cnr.it,

mariachiara.spadaro@unict.it

**Abstract**

Low-cost and environmentally friendly electrochemical energy storage systems are crucial to address the increasing global energy demand. Nanomaterials can play a pivotal role in catalysing charge storage and/or exchange, still the underlying mechanism often remains poorly investigated, as for ZnO/ZnS nanostructures onto Ni foam. In this work, we investigate hydrothermally grown ZnO/ZnS nanostructures decorating Ni foam for energy storage application. Morphology, structure and composition are evaluated via electron microscopy-based methodologies. The electrochemical energy storage performance is evaluated by cyclic voltammetry (CV) measurements with the aim to highlight the energy storage mechanism. When nickel foam (NF) is used as substrate, the system shows a predominant pseudocapacitive behaviour. By contrast, a modest and capacitive performance is measured on graphene paper (GP). Mott-Schottky (M-S) and open circuit potential (OCP) measurements suggests a key role of hole reservoir in ZnS decoration which boosts NF performances.

## Introduction

Nowadays the global energy consumption has dramatically increased[1], demanding low-cost and environmentally friendly devices able to store and deliver both high energy and power density, while sustaining long cycle life[2]. In this scenario, electrochemical energy storage systems, including batteries and supercapacitors, play a crucial role being able to store charge either by Faradaic process (batteries) or by capacitive mechanism (electric double-layer capacitors, EDLCs), or by a combination of the two (pseudocapacitors)[3]. Supercapacitors (EDLCs and pseudocapacitors) can safely provide high power and rapid charging, but with energy density significantly lower than batteries. Thus, extensive research is being conducted to develop energy storage systems that can replicate the storage capability of batteries, while preserving high power density and cycling stability[4,5]. The selection/optimization of the electrode material is the most critical parameter that tune the efficiency of these devices. Transition metal oxides (TMOs) are very promising[6] and, among them, zinc oxide (ZnO) has attracted significant attention owing to its excellent chemical and physical properties, being extremely abundant on Earth and non-toxic. ZnO can be easily synthesized in various and controllable morphologies by green, scalable and low cost chemical routes[7,8,9,10,11]. Various strategies have been applied to enhance the energy storage capacity of ZnO, including heterostructures[12], i.e. with ZnS. The latter possesses excellent theoretical capacity, high electrical conductivity, and strong redox activity, anticipating higher electrochemical performance than its oxide counterpart[13]. In this direction, Shah et.al.[14] fabricated ZnO/ZnS nanocomposites on nickel foam (NF), displaying a specific capacitance of 440.6 F/g at the current density of 1 A/g. Cao et.al.[15] testing ZnO/ZnS arrays on NF achieve a specific capacitance of 227 F/g at the current density of 1.28 A/g, and more studies[16,17].

When dealing with electrochemical devices the substrate plays a key role in the overall reaction [18]. One of the most widely used substrates for energy storage evaluation is NF[19,20,21,22,23], thanks to its high porosity, promising electrical and thermal conductivity, small ionic diffusion resistance, good corrosion resistance, and increased areal loading of active materials. Therefore, NF is the ideal substrate for highly stable electrodes with maximized contact areas with the active materials, which is beneficial for improved performance[24]. However, when using NF as substrate it is important to recall that bare NF is generally active in the charge storage mechanism[25], leading to an overestimation of the electrochemical performance and, therefore, the calculated specific capacitance[26]. Comparative studies conducted on different substrates have shown that the NF-supported active materials displayed much higher electrochemical performance than other substrate-supported material, such as Ti gauze[26], platinum or glassy carbon electrodes[25]. Therefore, one could easily question if the electrochemical performance evaluated can be really ascribed to the material or to its synergy with the NF.

To the best of our knowledge, literature studies[14,27] focus on the electrochemical performance evaluation of the entire system, without providing an in-depth analysis of the specific contribution of the substrate nor to its synergy with the investigated material.

In this work we aim in filling this gap proposing in depth investigation of ZnO and ZnO/ZnS nanostructure-based electrodes on NF for energy storage applications. Considering the crucial role of the substrate on the evaluation of electrochemical performance, the aim of this work is to investigate the ZnO/NF and ZnO/ZnS/NF systems and unveil the key players in electrochemical storage mechanisms. This study and our approach could be extended to various material/substrate combinations, going in the direction of unquestionably optimised energy storage devices to face next energy transition process.

## Experimental section

*Synthesis of ZnO and ZnO/ZnS nanostructures*

The nanostructures of interest were prepared starting from ZnO nanorods that were synthesized by the low-cost and environmentally friendly chemical bath deposition (CBD), placing an aqueous solution of 25 mM zinc nitrate hexahydrate ($Zn(NO_3)_2 \cdot 6H_2O$, Sigma-Aldrich, purum p.a., crystallized, ≥99.0%) and 50 mM hexamethylenetetramine ($C_6H_{12}N_4$, Sigma-Aldrich, puriss. p.a., reag. Ph. Eur., ≥99.5%) in a bain-marie configuration for 1 hour at 90°C. After washing and drying, the obtained ZnO powders were grounded in an agate mortar for about 5 minutes.

In the second step, ZnS surface rich nanostructures were obtained by exploiting the conversion of ZnO to ZnS by sulfidation. ZnO nanorods powder and thiourea ($NH_2CSNH_2$, Sigma-Aldrich, ACS reagent, ≥99.0%), with a molar ratio of thiourea/ZnO of 2.5, were dissolved in 15 ml of deionized water (Milli-Q, 18 MΩ·cm) and magnetically stirred for 30 minutes. After that, the solution was transferred to a 25 ml Teflon liner, inserted in an autoclave and placed in a laboratory oven for 24 hours at 200°C. After cooling to room temperature, the obtained solution was centrifuged at 6000 rpm, washed with deionized water and ethanol twice and dried on a hot plate at 100°C for about 3 hours.

*Electrode preparation*

Ni foam (NF) substrates (Goodfellow Inc., Huntingdon, England; thickness 1.6 mm and porosity ≥95%) were rinsed with deionized water and ethanol and dried with $N_2$. A small portion (3 mg) of ZnO and ZnS-based nanostructures' powder was dispersed in 1 mL of deionized water and, after 30 minutes of sonication, 75 µL of a solution of polyvinylidene difluoride (PVDF) in acetone (11 g/L) was added as binder. The mixture was further sonicated for 30 minutes. The nanostructures' dispersion was then drop cast on 1 $cm^2$ of NF substrate to obtain the electrodes. After the nanostructure loading, the electrodes were placed on a hot plate at 75°C to promote and accelerate the solvent evaporation. The deposited mass was calculated as the difference between the electrode mass after and before the drop-casting, by using a Mettler Toledo MX5 microbalance (Columbus, OH, USA) with a sensitivity of 0.01 mg. The measured mass is about 1 mg/$cm^2$. ZnO and ZnS-based electrodes were also prepared using graphene paper (GP) substrates (240 µm thick, Sigma Aldrich, St. Louis, MO, USA). The GP substrates were rinsed with deionised water and ethanol, followed by drying under $N_2$ flow. The same nanostructures' dispersion used for the NF substrates was drop cast on 1 $cm^2$ of GP substrate to obtain these electrodes.

*Characterization*

The morphology of the obtained nanostructures was analysed by using a scanning electron microscope (Gemini field emission SEM Carl Zeiss SUPRA 25, Carl Zeiss Microscopy GmbH, Jena, Germany).

The atomic structure was examined by high-resolution transmission electron microscopy (HRTEM) using a field emission gun FEI Tecnai F20 microscope. High-angle annular dark-field (HAADF) scanning transmission electron microscopy (STEM) was associated with electron energy loss spectroscopy (EELS) in a Tecnai microscope by using a GATAN QUANTUM energy filter to obtain chemical compositional maps. STEM-EELS maps were performed using O K-edge at 532 eV (green), Zn L-edge at 1020 eV (red), and S K-edge at 2472 eV (light-blue).

The electrochemical measurements were performed at room temperature under alkaline conditions, using 1M KOH as supporting electrolyte. Cyclic voltammetry (CV) measurements were carried out by using an Autolab PGSTAT204 potentiostat (Metrohm, Utrecht, The Netherlands) in a three-electrode setup with a platinum wire as counter electrode, a saturated calomel electrode (SCE) as reference electrode, and the samples deposited on NF and on GP substrates as working electrodes. CV curves were recorded at a scan rate of 20 mV/s scanning the potential from 0 to 0.45 V vs SCE.

Mott–Schottky (M–S) analyses were conducted using a VersaSTAT 4 potentiostat (Princeton Applied Research, Oak Ridge, TN, USA) in the potential range −0.65 ÷ 0.35 V vs Ag/AgCl at 1000 Hz frequency. Open circuit potential (OCP) measurements were also performed in dark and light conditions, using a multi-wavelength fiber coupled LED system in the UV range (375-405 nm).

## Results and discussion

The nanostructures (NSs) morphology can be seen in Figure 1. The SEM image of ZnO NSs (Figure 1a) reveals that most of the NSs appear in ZnO rods' clusters, recalling sea-urchins with smooth surface. The sulfidation process introduces spongy ZnS NSs on the ZnO surface (Figure 1b)[28,29,30]. Figure 1c shows the morphology of an electrode prepared by depositing a mixture of ZnS-based powders and polyvinylidene difluoride (PVDF, serving as binder) onto NF.

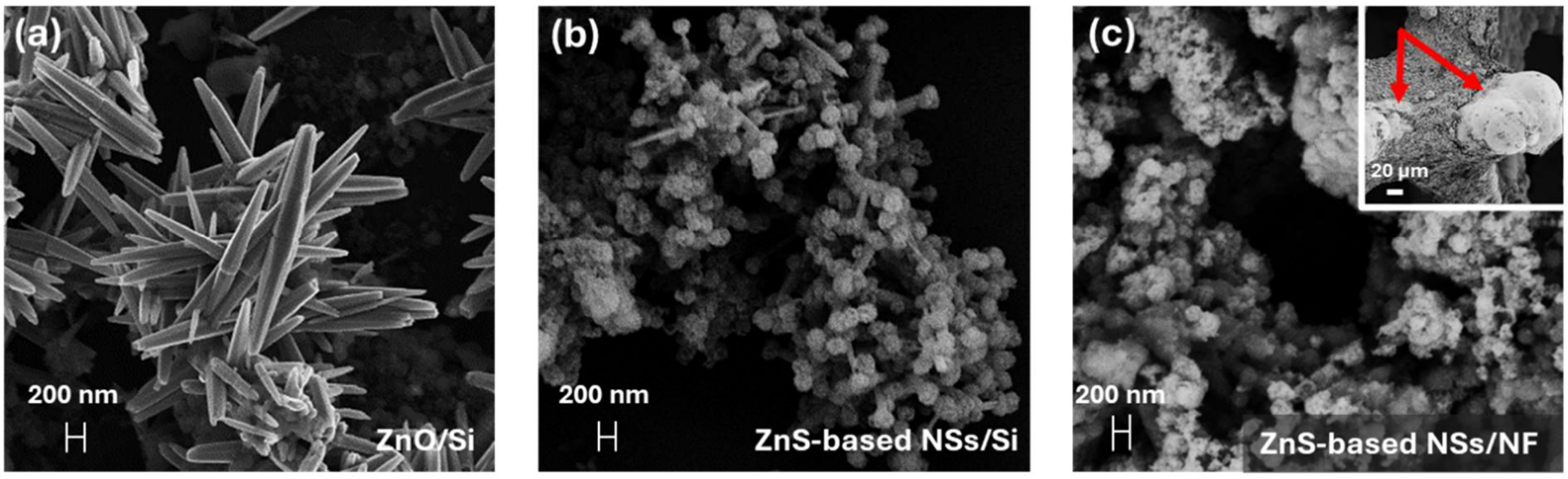

**Figure 1.** SEM images of (a) the ZnO nanorods on Si and (b) NSs obtained after the sulfidation process on Si. (c) ZnS-based electrode deposited onto NF (low-magnification SEM in inset). The inset displays a wider view of the decorated substrate, where the presence of exposed NF regions, marked by the red arrows, indicates an uneven surface coverage.

To investigate the structural changes induced by sulfidation in ZnO nanorods, TEM and HRTEM analyses were performed (Figure 2). The ZnO nanorods HRTEM image and the corresponding power spectrum analysis (Figure 2c) clearly show the hexagonal wurtzite phase, belonging to the P63/mc (186) space group. The power spectrum analysis of the ZnS-based NSs (Figure 2d) shows a cubic sphalerite phase, belonging to the F4-3m (216) space group. In addition, STEM-EELS maps reported in Figure 2e show that zinc and oxygen are homogeneously distributed within the rods. After the sulfidation process, the spongy nanoparticles do not present oxygen traces confirming that they have been homogeneously sulfided (Figure 2f), while in some minor area oxygen signal was detected belonging to the ZnO nanorods that have not been modified during sulfidation (Figure 2g); we therefore refer to the sulfided material as ZnO/ZnS.

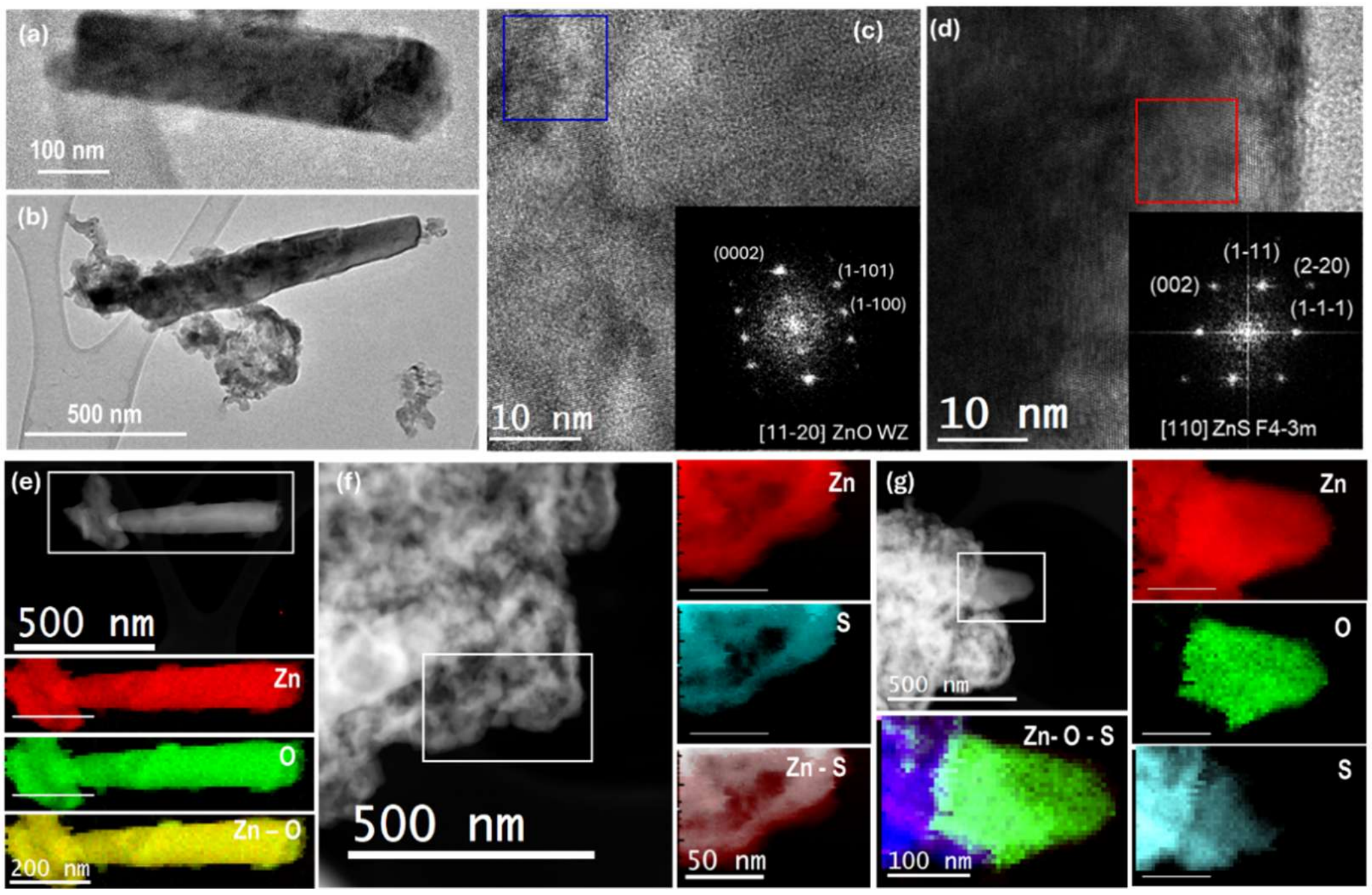

**Figure 2.** Low-magnification TEM (a,b), HRTEM (c,d) images and the corresponding indexed power spectra of ZnO and ZnO/ZnS, respectively. HAADF-STEM and STEM-EELS maps of the region highlighted in white box of ZnO (e) and ZnO/ZnS (f,g). STEM-EELS maps were performed using O K-edge at 532 eV (green), Zn L-edge at 1020 eV (red), and S K-edge at 2472 eV (light-blue).

The material’s electrochemical performance was evaluated through CV measurements and compared with bare NF (Figure 3a). As expected, the bare substrate is electrochemically active, exhibiting a pseudocapacitive behaviour attributed to the Faradic reactions:

$$\mathrm{Ni(OH)_2 + OH^- \leftrightarrow NiOOH + H_2O + e^-}$$

resulting in a pair of redox peaks corresponding to the reversible transition $\mathrm{Ni^{+2}/Ni^{+3}}$ in alkaline medium[26,31,32,33]. When the number of CV cycles increases, we expect a shift of the anodic and cathodic peaks to higher potentials, due to a phase transformation of nickel hydroxide and nickel oxyhydroxide, together with an increase in the peak area[25]. Therefore, the prepared electrodes underwent the same number of cycles. As shown in Figure 3a, the CV curves of ZnO and ZnO/ZnS on NF also reveal the presence of both anodic and cathodic peaks, in agreement with previous literature’s reports for similar electrodes[14,27]. The high porosity of NF ensures a well-defined dispersion of Zn-based NSs on NF surface, while maintaining a high portion of its surface exposed (red arrows in Figure 1c). Therefore, it would not be surprising to associate the observed peaks to the redox reactions of nickel, potentially influenced by the presence of the deposited material. In fact, when the NF is coated with the active material, the potential drop (mainly caused by the thickness of the active material) is expected to increase, requiring a higher potential for the redox reaction to occur, as observed in Figure 3a, where the redox peaks shift towards higher potentials for ZnO and ZnO/ZnS on NF.

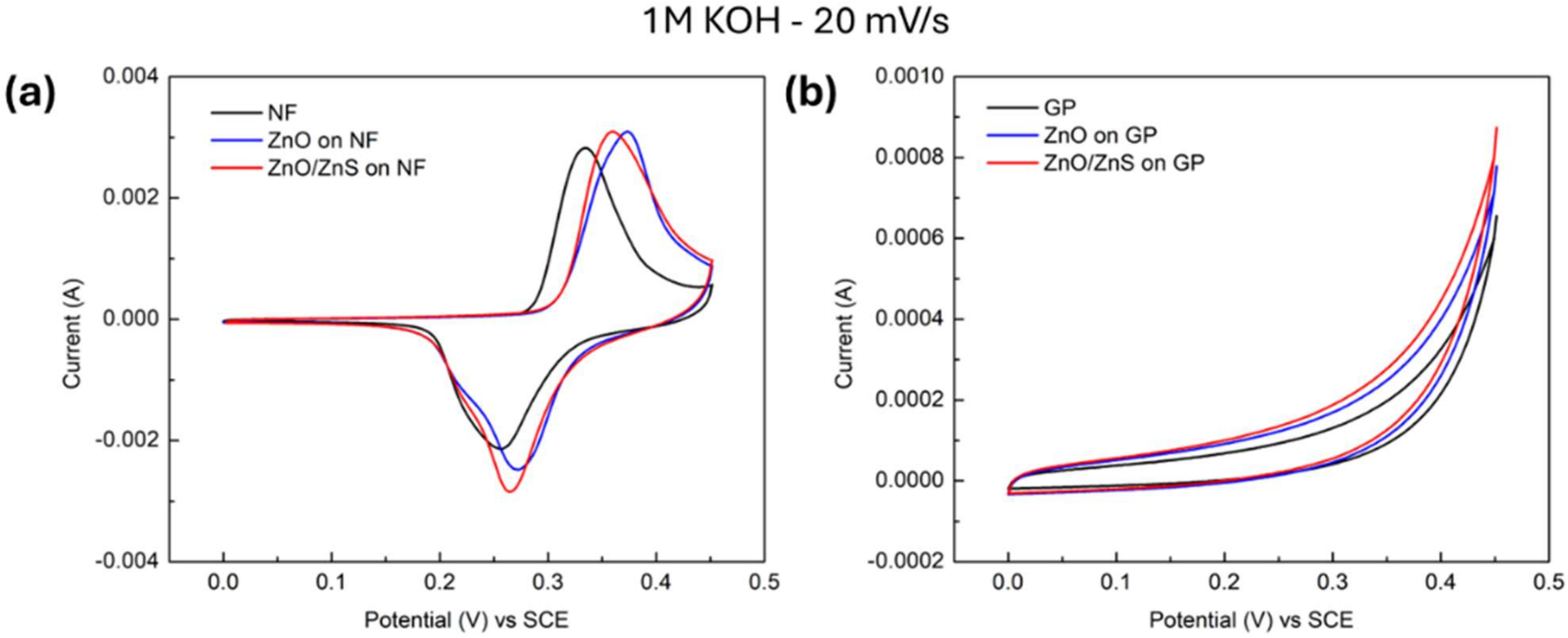

**Figure 3.** CV curves in 1M KOH at 20 mV/s in the potential range 0-0.45 V vs SCE of the electrodes deposited on (a) NF and (b) GP.

To properly evaluate the electrochemical contribution of the NSs alone, it is important to isolate their response from the substrate[7]. Therefore, within the identical potential range and by using the same supporting electrolyte, the investigation has been performed also using a nickel-free substrate, such as graphene paper (GP) (Figure 3b). For all the electrodes, the CV measurements show similar trend to GP, which is known to behave like EDLC[34]. Hence, the anodic and cathodic peaks, characteristic of the CVs of the electrodes on NF, cannot be associated with faradic redox reactions of Zn-based material in alkaline medium[14,16,27,31] but originate primarily from NF itself. This observation, corroborated by the shape of CV curve obtained with Zn-based electrode on GP (Figure 3b), suggests that the investigated material predominantly exhibits a capacitive behaviour. To quantitatively evaluate the performance of the different samples, the stored charge Q (C) was determined from the CV curves as follows[35,36]:

$$Q = \frac{\int IdV}{\nu}$$

where $V$ is the applied potential (V), $I$ is the measured current (A) and $\nu$ is the scan rate (V/s). For a robust evaluation of the NSs' performance and to avoid misinterpretation of the measured effect[26], the contribution of the bare substrate must be accounted and subtracted from the total stored charge of the full electrode (especially when using NF). Table 1 summarizes the value of the stored charge (Q) of the bare NF as well as of the electrodes consisting of ZnO on NF and ZnO/ZnS on NF. ΔQ represents the net charge stored by the NSs and it is defined as the difference between the charge stored by the complete electrode and the one of the bare substrate. The relative increment is calculated as the ratio ΔQ/Q. When the NF is covered with ZnO NSs, ΔQ/Q increases to 17%. NF-supported ZnO/ZnS electrode shows a higher increment, approximately 25%, providing a clear indication of the higher charge storage performance of the ZnS-based electrode. The net specific capacitance $C_s$ (F/g) of the ZnO and ZnO/ZnS NSs was calculated by considering the net stored charge (ΔQ), using the following[7,37]:

$$C_s = \frac{\Delta Q}{m\Delta V}$$

where $m$ is the active mass (g) and ΔV is the potential range (V), and results approximately 9 F/g and 13 F/g for ZnO and ZnO/ZnS, respectively.

| Sample | Q (mC) | ΔQ/Q (%) | $C_s$ (F/g) |
|---|---|---|---|
| NF | 22.1 | - | - |
| ZnO on NF | 25.9 | 17 | 9 |
| ZnO/ZnS on NF | 27.7 | 25 | 13 |

**Table 1.** Stored charge of bare substrate, ZnO and ZnO/ZnS, percentage increase compared to bare substrate and specific capacitance of ZnO and ZnO/ZnS on NF.

Similar calculations were conducted for the electrodes on GP, ensuring a consistent comparison. The analysis, performed from the CV curves shown in Figure 3b, indicates that when the GP substrate is coated with ZnO and ZnO/ZnS, the amount of charge involved in the process increases to 1.1 mC and 1.8 mC, respectively, corresponding to an increment of 23% and 38% compared to bare GP. The net storage capacity allows the determination of an effective specific capacitance of 13 F/g and 20 F/g for ZnO and ZnO/ZnS, respectively. It should be noted, however, that the potential range used for the CV measurements was not optimized for this type of electrode on GP, but still the results highlight an enhanced charge absorption capability of ZnO/ZnS NSs compared to ZnO, in line with the observations for electrodes on NF. The stored charge increment, compared to the bare substrate, appears to be higher on GP than on NF, suggesting a substrate dependence's storage efficiency. Smooth and homogeneous substrates, such as GP, improve the accessibility of the active surface, facilitating the charge accumulation at the electrode-electrolyte interface, thereby promoting capacitive processes which are surface-controlled. In contrast, the porous and rough morphology of NF may partially suppress purely capacitive charge storage. This interpretation is consistent with the higher $C_s$ obtained for GP supported electrodes compared to their NF counterparts. Overall, switching from ZnO to ZnO/ZnS leads to improved electrochemical performance: for GP supported electrodes, this enhancement can be attributed to the more capacitive nature of ZnS, favouring surface-controlled charge storage. Interestingly, although the capacitive contribution is less dominant on NF, an increase in electrochemical performance is still achieved. The latter suggests that ZnS plays a different role on NF: it is proposed to provide holes that catalyse Ni redox reactions, rather than promoting charge storage itself.

To get a deeper insight into the electrochemical behaviour and investigate in more detail the combination of ZnO/NF and ZnO/ZnS/NF, M-S measurements are reported in Figure 4. As the potential increases, $1/C^2$ values go to zero evidencing a negative slope of the linear region, typical for p-type semiconductor[38,39]. It is worth noting that all the electrodes exhibit a p-type behaviour owing to the predominant contribution of NF[40], as already mentioned.

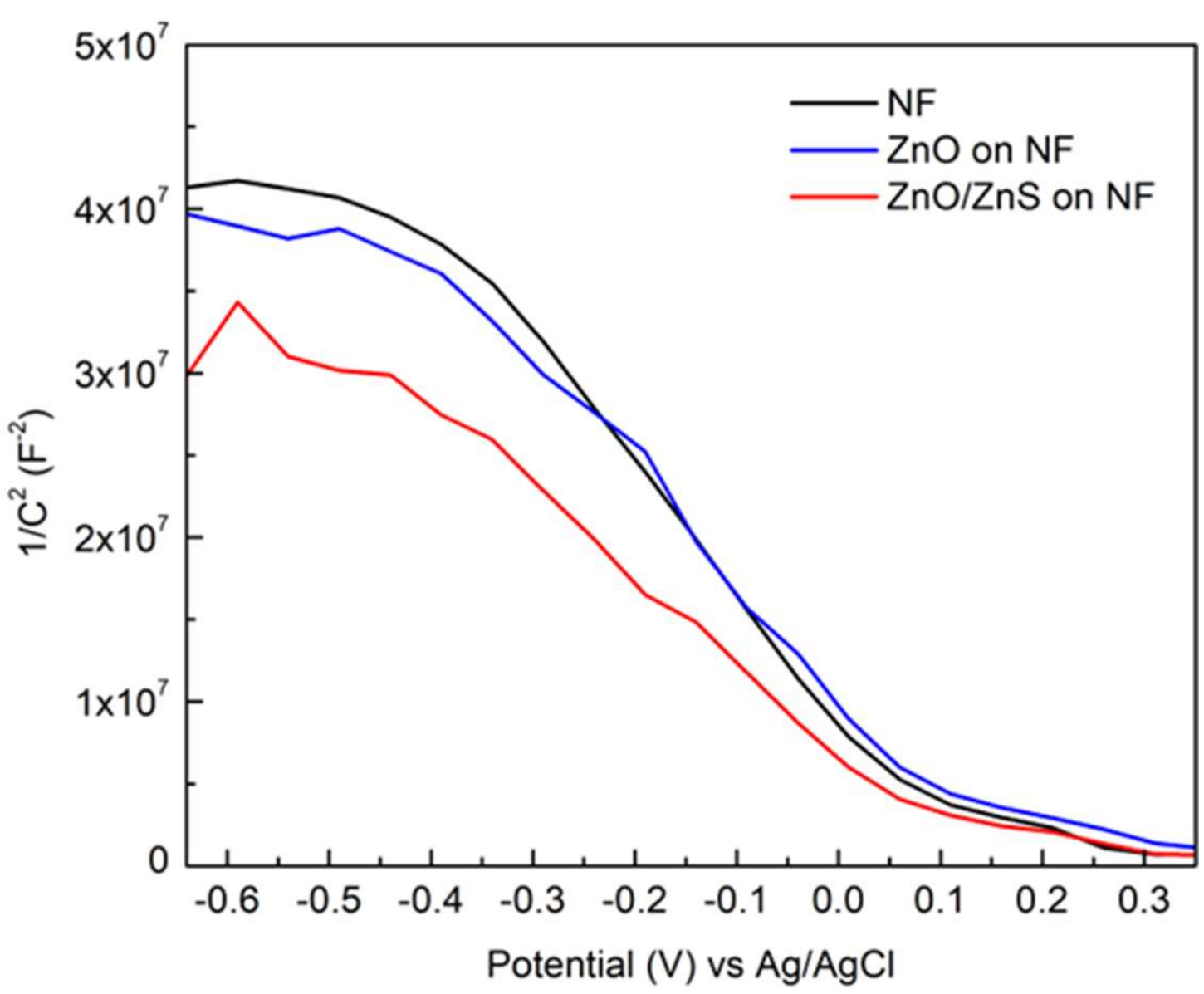

**Figure 4.** Mott-Schottky plot of bare substrate, ZnO and ZnO/ZnS.M-S measurements were performed in 1M KOH aqueous solution. The inverse square of the capacitance, measured as a function of potential applied to the sample under study, is here reported.

Using the M-S plot it is possible to determine fundamental properties of semiconductor-electrolyte systems: (i) the band bending (B.B.) resulting from the alignment of the Fermi level of the semiconductor with the Fermi level of the electrolyte under equilibrium conditions[41], (ii) the flat band potential ($E_{FB}$) that is applied if no band bending at the semiconductor-electrolyte interface would occur[42]. $E_{FB}$ was determined from the x-axis intercept of the linear region of the plot according to the relation[39]

$$\frac{1}{C^2} = \frac{-2}{\varepsilon_0 \varepsilon_r e N_A}\left(E - E_{FB} - \frac{kT}{e}\right)$$

where $C$ represents the space region capacitance, $\varepsilon_0$ is the permittivity of vacuum, $\varepsilon_r$ is the dielectric constant of the semiconductor, $e$ is the electron charge, $N_A$ denotes the density of acceptors, $E$ is the applied potential, $E_{FB}$ represents the flat band potential, $k$ is the Boltzmann constant and $T$ is the temperature in Kelvin. $E_{FB}$ is related to band bending at the liquid-semiconductor interface under equilibrium conditions. Therefore, by combining this value with the open-circuit potential ($E_{OC}$), recorded at the beginning of each M-S analysis, the actual band bending can be defined for each electrode as the difference between $E_{FB}$ and $E_{OC}$[43,44,45] (Table 2).

| Sample | $E_{FB}$ (mV) vs Ag/AgCl | $E_{OC}$ (mV) vs Ag/AgCl | B.B. (mV) vs Ag/AgCl |
|---|---|---|---|
| NF | 80 ± 2 | -277 ± 2 | 358 ± 3 |
| ZnO on NF | 118 ± 2 | -175 ± 2 | 293 ± 3 |
| ZnO/ZnS on NF | 98 ± 2 | -206 ± 2 | 303 ± 3 |

**Table 2.** Values of flat band potential, open circuit potential and band bending. The uncertainty on $E_{FB}$ and $E_{OC}$ was obtained by considering the applied voltage accuracy of the instrument (± 0.2% of value ± 2 mV), while the uncertainty on B.B. was calculated through the error propagation.

It can be immediately observed that the band bending at the equilibrium is more pronounced for bare NF if compared to the decorated ones. This implies a stronger electric field in the NF, leading to a higher accumulation of positive charges at the solid-electrolyte interface, promoting the accumulation of $OH^-$ ions on the electrolyte side and favouring the oxidation reaction of NF. In fact, the oxidation peak of bare NF appears at lower potentials when compared to the Zn-based electrodes (showing a reduced band bending), as observed in Figure 3a and in literature[43].

Further electrochemical analyses were carried out to elucidate the Zn-based material's contribution. OCP measurements' results upon dark and UV illumination (375 and 405 nm) are reported in Figure 5, after stabilising the electrodes. For bare NF OCP value remains unchanged upon illumination (Figure 5a). When NF is covered with ZnO NSs, OCP varies by approximately 5 mV upon illumination (Figure 5b), while with ZnO/ZnS NSs it undergoes to variation of approximately 20 mV (Figure 5c), showing a greater dark-light OCP variation.

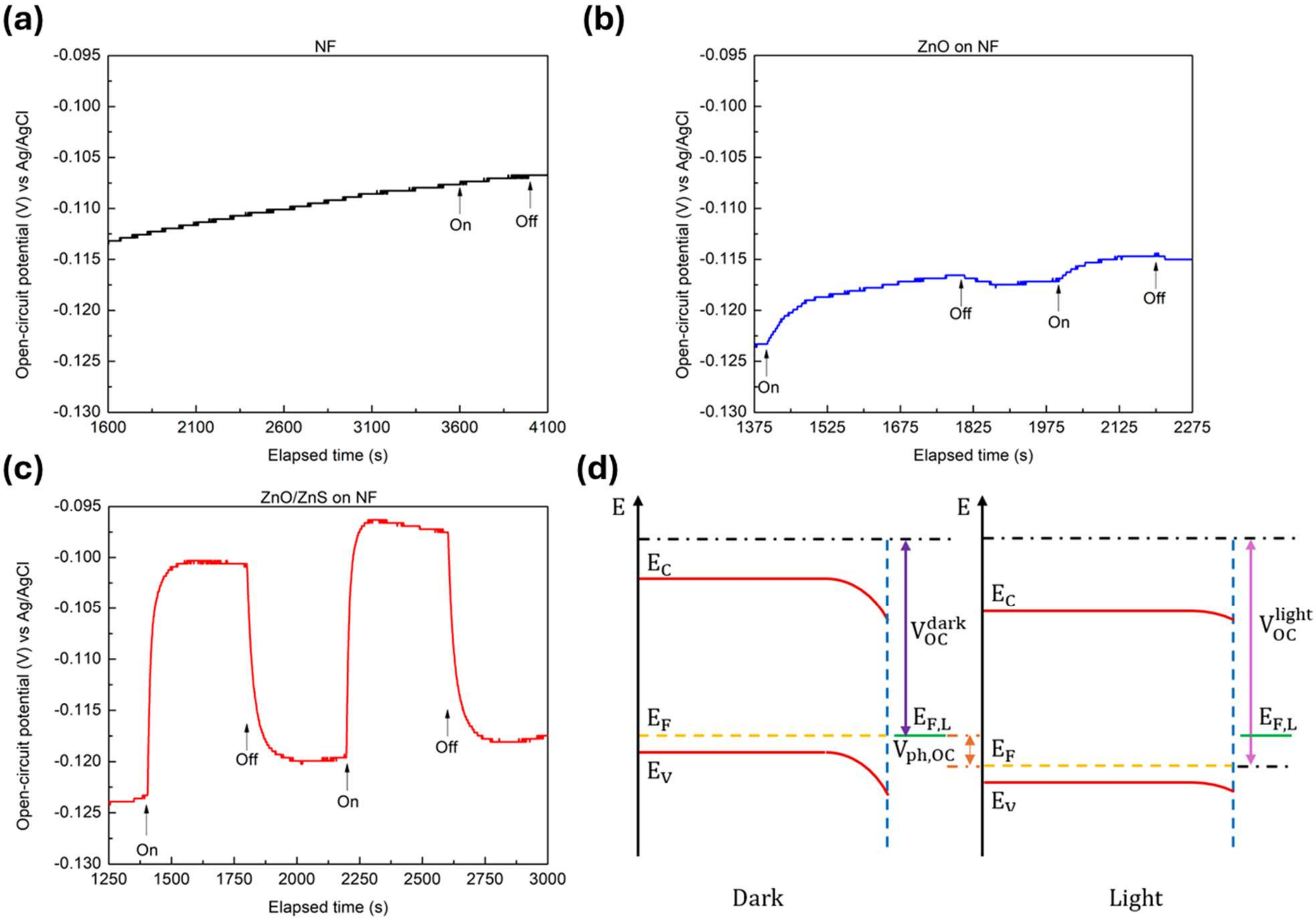

**Figure 5.** Open-circuit potential measurements in dark and light conditions of (a) bare substrate, (b) ZnO and (c) ZnO/ZnS. (d) Schematic of open-circuit energy band diagram under dark condition (left panel) and under light irradiation (right panel).

It is well known that the OCP shift upon illumination indicates the conductivity type of materials[46]. Under dark condition and at equilibrium, the Fermi level of the semiconductor ($E_F$) is aligned with the one of the electrolyte ($E_{F,L}$), generating a built-in potential that results in a downward band bending for p-type semiconductors[38] (Figure 5d, left panel). The OCP in dark, measured between the reference electrode and the working electrode, is given by $V_{OC}^{dark}$[47], as shown in Figure 5d. Under light irradiation, electron-hole pairs are generated and separated by the electric field within the space charge region[46]. The minority charge carriers (photogenerated electrons) move towards the semiconductor-electrolyte interface, while the majority charge carriers (holes) are driven toward the bulk of the semiconductor. The latter generates an additional electric field, in the opposite direction to the electric field of the space charge region, that can partially compensate the negative charges, reducing the band bending (Figure 5d, right panel). The OCP under illumination, measured between the reference electrode and the working electrode, is given by $V_{OC}^{light}$, as shown in Figure 5d. Its difference with respect to $V_{OC}^{dark}$ is ascribed to the generated photovoltage electrostatic potential $V_{ph,OC}$[47]. For a p-type semiconductor $V_{OC}^{light} = V_{OC}^{dark} + V_{ph,OC}$, indicating that OCP values are higher in the transition from dark to light if the material is p-type; while OCP decreases with illumination if the material is n-type. Therefore, evaluating the difference in OCP for each electrode upon illumination, we expect to find a higher value for electrodes with a noticeable p-type behaviour, as observed for ZnO/ZnS NSs when compared with ZnO alone, that tends to be n-type.

We can now elucidate the CVs trend (Figure 3a) and the increased stored charge with ZnS-based NSs. ZnO/ZnS material may act as reservoir of holes, being more p-type, as also confirmed by the lower slope of the linear region in the M-S plot, linked to a higher charge carrier density. The latter is expected to (i) promote the capacitive absorption of anions (from the electrolyte dissociation) and to (ii) catalyse the oxidation reaction of NF, ascribed to the synergy of NF with ZnO and even more with ZnO/ZnS NSs (exhibiting higher charge storage performance).

Furthermore, OCP measurements upon illumination evidence that the time required for band equilibrium is longer in ZnO NSs compared to ZnO/ZnS. The latter could be linked to a higher density of surface trap states in ZnO, which could hinder the separation of photogenerated charge carriers, resulting in a smaller OCP variation. The presence of such trap states provides insights into the electrochemical behaviour of the electrodes and offers a plausible explanation for the enhanced charge storage capacity of ZnO/ZnS. In this context, it is reasonable to assume that a higher number of charges remains confined within the trap states over extended timescales in ZnO NSs compared to ZnO/ZnS, thus limiting their participation in electrochemical processes and reducing ZnO charge storage capacity.

Our results highlight the role Zn-based NSs, especially when combined with Ni, emphasizing the importance of evaluating the individual contribution of each material to better monitor and understand the effect of the single constituents in the overall electrochemical process. Our findings pave the way for the rational design of optimised nanostructured electrodes for electrochemical energy storage.

## Conclusions

Hydrothermally grown ZnO/ZnS NSs are investigated for energy storage, by covering NF with ZnO and ZnO/ZnS nanomaterials. Bare NF is electrochemically active with a pseudocapacitive behaviour. NF contribution remains predominant even when covered with the active material, showing redox peaks that disappear when the NSs are deposited onto a nickel-free substrate, such as GP. However, the area under the CV curves shows an increment for ZnO and ZnO/ZnS NSs compared to bare NF (17% and 25% respectively). To elucidate this effect, an in-depth study of the ZnO/NF and ZnO/ZnS/NF combination was conducted via M-S measurements, showing that all electrodes exhibit a p-type behaviour. OCP measurements, under dark and light conditions, reveal that ZnO/ZnS NSs exhibit a larger variation compared to ZnO NSs, suggesting a more pronounced p-type behaviour. We then conclude that ZnO/ZnS on NF is a stronger reservoir of holes, both promoting the absorption of anions in solution (capacitive mechanism) and catalysing the oxidation reaction of NF (redox process). These results and the presented methodology suggest the importance of evaluating the effect of each constituent in a complex system such as ZnO/ZnS/NF combination, enabling to optimise the preparation of electrodes for energy storage applications.

## Acknowledgments

The authors acknowledge Giuseppe Pantè and Carmelo Percolla (CNR-IMM) for the assistance during the experiments. This work was funded by the European Union – next Generation EU, Mission 4, Component 2 via the project NanoStore, under the "PRIN PNRR 2022" call, project ID: P20227TNLX, CUP numbers: I53D23006650001 and B53D23028380001. This work was supported by the project NANO STRENGTH - Linea di Intervento 1- Progetti di Ricerca Collaborativa del "PIAno di inCEntivi per la RIcerca di Ateneo 2024/2026 Università di Catania. EU HORIZON-INFRA-2021-SERV-01 GA number 101058414 (ReMADE@ARI). Authors acknowledge the Joint Electron Microscopy Center at ALBA (JEMCA) and Grant IU16-014206 (METCAM-FIB) funded by the EU European Regional Development Fund (ERDF) and Generalitat de Catalunya 2021SGR00457. ICN2 is founding member of e-DREAM[48] and supported by the Severo Ochoa program from Spanish MCIN/AEI (Grant No.: CEX2021-001214-S) and is funded by the CERCA Programme/Generalitat de Catalunya.